\documentclass[review,onefignum,onetabnum]{siamart190516}

\usepackage{lipsum}
\usepackage{amsfonts}
\usepackage{graphicx}
\usepackage{epstopdf}
\usepackage{algorithmic}
\ifpdf
  \DeclareGraphicsExtensions{.eps,.pdf,.png,.jpg}
\else
  \DeclareGraphicsExtensions{.eps}
\fi

\newsiamremark{remark}{Remark}
\newsiamremark{hypothesis}{Hypothesis}
\crefname{hypothesis}{Hypothesis}{Hypotheses}
\newsiamthm{claim}{Claim}

\headers{MuyGPs}{A. Muyskens, B. Priest, I. Goumiri, and M. Schneider}

\title{MuyGPs: Scalable Gaussian Process Hyperparameter Estimation Using Local Cross-Validation\thanks{Submitted to the editors April 29, 2021.
\funding{This work was performed under the auspices of the U.S. Department of Energy by Lawrence Livermore National Laboratory under Contract DE-AC52-07NA27344 with IM release number LLNL-JRNL-822013.
Funding for this work was provided by LLNL Laboratory Directed Research and Development grant 19-SI-004.}}}

\author{Amanda Muyskens\thanks{Engineering Division, Lawrence Livermore National Laboratory\\
Livermore, CA 94550, USA
   (\email{muyskens1@llnl.gov}).}
\and Benjamin Priest\thanks{Center for Applied Scientific Computing, Lawrence Livermore National Laboratory\\
Livermore, CA 94550, USA}
  \and Im\`ene Goumiri\thanks{Physics Division, Lawrence Livermore National Laboratory\\
Livermore, CA 94550, USA}
\and Michael Schneider\thanks{Physics Division, Lawrence Livermore National Laboratory\\
Livermore, CA 94550, USA}}

\usepackage{amsopn}

\ifpdf
\hypersetup{
  pdftitle={MuyGPs},
  pdfauthor={A. Muyskens, B. Priest, I. Goumiri, and M. Schneider}
}
\fi

\usepackage{subcaption}

\begin{document}

\maketitle

\begin{abstract}
Gaussian processes (GPs) are non-linear probabilistic models popular in many applications.
However, na\"ive GP realizations require quadratic memory to store the covariance matrix and cubic computation to perform inference or evaluate the likelihood function.
These bottlenecks have driven much investment in the development of approximate GP alternatives that scale to the large data sizes common in modern data-driven applications.
We present in this manuscript \texttt{MuyGPs}, a novel efficient GP hyperparameter estimation method.
\texttt{MuyGPs} builds upon prior methods that take advantage of the nearest neighbors structure of the data, and uses leave-one-out cross-validation to optimize covariance (kernel) hyperparameters without realizing a possibly expensive likelihood.
We describe our model and methods in detail, and compare our implementations against the state-of-the-art competitors in a benchmark spatial statistics problem.
We show that our method outperforms all known competitors both in terms of time-to-solution and the root mean squared error of the predictions.  
\end{abstract}



\section{Introduction}
\label{sec:intro}

Gaussian process (GP) models are widely popular nonlinear models ubiquitous in spatial statistics \cite{gelfand2010handbook, stein2012interpolation} and commonly utilized in machine learning applications \cite{rasmussen2006gaussian, liu2020gaussian}.
Gaussian process regression (GPR) tractably accounts for the correlation among all observed data, making them favored models in the interpolation of highly-nonlinear responses.
GPR is popular in practice due to its sample efficiency and closed-form posterior distribution inference.
Consequently, GPs are widely employed in applications where observations are expensive to record or simulate and where overconfident prediction or extrapolation has a high cost.

A process is said to follow a GP if any finite set of $n$ realizations of the process follows a multivariate normal distribution.
We conventionally assume that the covariance has a parametric form depending on a pairwise \emph{kernel function} $K_\theta(\cdot, \cdot)$ depending upon hyperparameters $\theta$.
Unfortunately, learning $\theta$ is typically difficult and computationally expensive.
The computing and storing a covariance has $O(n^2)$ cost, while realizations of GPR and evaluating the likelihood required in the conventional training of $\theta$ have $O(n^3)$ cost.
Consequently, training a full GP model is prohibitively expensive for large $n$ using common hardware.

The design of computationally efficient methods of Gaussian process estimation is an active area of research. 
Many approximate methods for GP inference attempt to sparsify the correlation matrix. 
Some approaches partition the domain into spatially contiguous partition blocks and estimate independent models within those blocks \cite{gramacy2007tgp, fuentes2002spectral, sang2011covariance}.
Alternately, others randomly sample data partitions across the domain for independent Bayesian stationary GP analyses and subsequently use the geometric median of the subset posteriors as a means of model averaging \cite{guhaniyogi2018meta}.
Covariance tapering \cite{furrer2006covariance, furrer2010spam} alters the covariance matrix so that near-zero entries are considered independent and therefore set to zero.
Finally, the locally approximate GP (laGP) \cite{gramacy2015local, gramacy2016lagp} uses a similar sparsification - although it does not realize an overall covariance matrix - by fitting independent models within a local window of each prediction location and assuming that all other observations.

Other methods assume instead that the precision matrix, the inverse of the covariance matrix, is sparse.
One approach is to assume that data observed on a regular grid is spatially autogressive, which yields a sparse precision matrix, and to employ multiple resolutions of radial basis functions for efficient prediction  \cite{nychka2015multiresolution}.
Next, \cite{lindgren2011explicit} rely on the equivalence between stochastic partial differential equations (PDEs) and the Mat\'ern covariance fields.
This method induces sparsity in the precision matrix by fitting piecewise linear basis functions on a triangularization of the domain.
Other approaches still rewrite the multivariate normal likelihood in terms of chained conditional distributions, inducing sparsity by modeling only a subset of conditioning sets, such as those induced by the nearest neighbors structure of the data \cite{vecchia1988estimation, datta2016hierarchical}.
This is mathematically equivalent to a sparse precision matrix defined through its Cholesky decomposition. 

Spectral methods form effective models on gridded data.
Such methods forgo modeling the correlation function in favor of the spectral density, which is the Fourier transformation of the covariance matrix.
This allows estimating $\theta$ via the Whittle Likelihood \cite{whittle1954stationary} or stochastic score approximation \cite{stein2013stochastic}. 
Although the Whittle likelihood requires a full grid, \cite{guinness2019spectral} use an iterative imputation scheme to allow for missing values. 
Further, \cite{muyskens2018non} demonstrate a fine grid can be used to approximately apply spectral methods to irregularly-spaced observations.

Variational low-rank approximations to the covariance are particularly popular in the machine learning literature \cite{lazaro2011variational, tran2015variational, gibbs2000variational}.
The spatial statistics literature refers to such models as predictive processes, where a small number of knot locations across the domain induce a low-rank Nystr\"om approximation to the dense covariance matrix \cite{banerjee2008gaussian}.
A related approach fits compactly supported basis functions on recursively partitioned subregions of the domain, using computations that can be parallelized \cite{katzfuss2017multi, jurek2019multiresolution}. 
This basis function selection uses a low-rank approach similar to that of the predictive process \cite{banerjee2008gaussian}. 
Another method that relies on multiple resolutions for estimation uses a spatial process assumed to be the sum of a set of resolutions of Gaussian basis functions, which they refer to as basic areal units (BAUs) \cite{zammit2017frk, zammitmangion2018frk}.

Other methods do not rely on Gaussian processes, or any statistical distribution at all.
For example, the Gapfill method relies on quantile regression within a local window of the observation \cite{gerber2018predicting} .
Since this method is originally described for space-time data, but the data is often over space only, the "time" dimension is approximated by shifting the original image \cite{heaton2019case}.

Heaton et al. \cite{heaton2019case} directly compared many of these methods on a benchmark land surface temperature dataset. 
This survey is particular notable because the authors coordinated a blind competition, where each team implemented the method in which they invented or otherwise owned expertise.
Therefore, there was no chance of a method underperforming due to misunderstanding or misuse of the model or method.
The central data problem on which the paper focuses is the prediction of missing measurements from a large land surface temperature gridded dataset sourced from the MODIS satellite.
Instead of sampling the testing set randomly across the domain, the authors use realistic cloud coverage from an image taken on another day to form a realistic and challengingly irregularly-shaped region of missing data.
Ultimately, the authors conclude that there is no one best method for computationally efficient GP estimation as some methods produce estimates quickly while others are more accurate in terms of root mean squared error (RMSE). 
Additionally, \cite{edwards2020precision} demonstrate their new methods on the same dataset.
These methods are improved in terms of time of estimation from most of the original methods, but are not the top competitors in terms of RMSE. 

\begin{figure}
	\includegraphics[width=\linewidth]{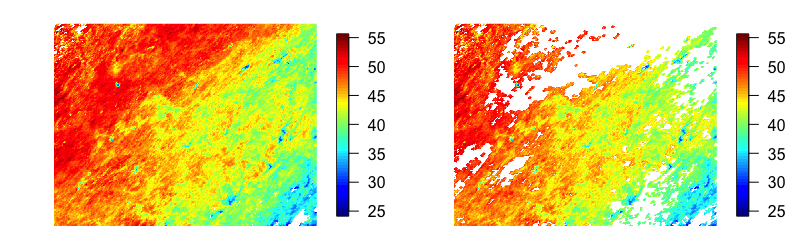}
	\caption{Full (left) and training (right) land surface temperature datasets used in \cite{heaton2019case} as a benchmark competition and sourced from the MODIS satellite on August 4, 2016.}
	\label{fig:data}
\end{figure}

We have developed a GP estimation method designed to remain both accurate and fast as data size grows.
We invoke sparsity through locality neighborhoods inspired by methods like \cite{gramacy2015local}.
However, instead of assuming independent models within those local neighborhoods, we assume an overall stationary GP model across the domain.
Further, we entirely avoid maximum likelihood evaluation in favor of leave-one-out cross-validation, where we produce predictions using only the local neighborhoods.
This amounts to the assumption that the weights that are applied to the data vector in order to obtain the predictions (kriging weights) are sparse, rather than the covariance matrix or precision matrix as in previous approximation methods.
We formally optimize the hyperparameter values against the leave-one-out cross-validation objective function.
We limit ourselves in this work to mean squared error loss for our objective functions, but other loss functions are possible.
We then use these estimated hyperparameters to realize responses at the prediction locations, which are similarly formulated with only their nearest neighbors.

Leave-one-out cross-validation is typically a computationally expensive process when predictions are obtained using the entire dataset.
In fact, it can be more expensive than the original maximum likelihood estimation problem ($O(n^4)$), since a large matrix ($(n-1)\times (n-1)$)must be inverted for each prediction. 
However, by limiting predictions to be based on only their $k$ nearest neighbors, computation of each prediction reduces to many small, parallelizable solves instead of one prohibitively large solve.
We can further improve our time-to-solution by relying upon a batch of $b$ training examples for the leave-one-out training procedure, reducing the overall time complexity to $O(bk^3)$.
In addition to making the time complexity formally independent of $n$, this dramatically improves cross validation speed when $b \ll n$ and $k \ll n$.
Additionally, obtaining the $k$ nearest neighbors is more efficient to compute than likelihood evaluation, and can be further accelerated by approximate KNN algorithms.
We find in our experiments that this deceptively simple method is effective, fast, and leaves open the opportunity for further future acceleration by way of distributed computation.

In this manuscript, we develop a novel method of approximate Gaussian processes hyperparameter estimation that is efficient, scalable and accurate.
In section \ref{sec:gp} we review a general stationary GP model.
In section \ref{sec:muygp} we describe our novel hyperparameter estimation method.
In section \ref{sec:sim} we demonstrate the performance of our method in the benchmark dataset from \cite{heaton2019case}, and show it performs favorably in accuracy and computation time in comparison to all of the existing state-of-the-art competitors.
Finally in \ref{sec:discuss}, we summarize our work and outline future advancements that could be made in order to improve on our methods.

\section{Background: Gaussian Processes}
\label{sec:gp}

We will consider throughout a univariate response $Y : \mathcal{X} \rightarrow \mathbb{R}$, where $\mathcal{X} \subseteq \mathbb{R}^p$ is the observation space.
For notational convenience and by convention we assume that $Y$ is de-trended and therefore has zero mean. 
Extensions to non-zero and multivariate processes are trivial, so we will avoid them for the sake of clarity.
We say that $Y$ follows a Gaussian process if the response at any finite set of $n$ points $X = (\mathbf{x}_1, \dots, \mathbf{x}_n) \in \mathcal{X}^n$ follows a multivariate normal distribution.
That is,
\begin{equation} \label{eq:gp_prior}
	Y(X) = (Y(\mathbf{x}_1), \dots, Y(\mathbf{x}_n))^T \sim \mathcal{N} \left ( \widetilde{0}, K_\theta(X, X) \right ),
\end{equation}
where $\mathcal{N}$ is the multivariate Gaussian distribution, $\widetilde{0}$ is the $n$-dimensional zero vector, and $K_\theta(X, X)$ is an $n \times n$ positive definite, symmetric covariance matrix between the elements of $X$ that is controlled non-linearly through kernel function $K_\theta(\cdot, \cdot)$ and hyperparameters $\theta$.

Similarly, any finite set of $n^*$ unobserved data $X^* = (\mathbf{x}^*_1, \dots, \mathbf{x}^*_{n^*}) \in \mathcal{X}^{n^*}$ is also jointly normal with observed data $X$ by the GP assumption.
Thus, the conditional distribution for the response at the new locations $X^*$ given responses observed at $X$ is also multivariate normal with mean and variance
\begin{align}
	\widehat{Y}(X^* \mid X) &= K_\theta(X^*, X) K_\theta(X, X)^{-1} Y(X), \text{ and}
	\label{predmean}\\
	\text{Var}(\widehat{Y}(X^* \mid X))  &= K_\theta(X^*, X^*) - K_\theta(X^*, X) K_\theta(X, X)^{-1} K_\theta(X, X^*),
	\label{predvar}
\end{align}
where $K_\theta(X^*, X) = K_\theta(X, X^*)^T$ is the cross-covariance matrix between the elements of $X^*$ and $X$, and $K_\theta(X^*, X^*)$ is the covariance matrix between the elements of $X^*$, similar to $K_\theta(X, X)$.

Note that the conditional mean in Equation~\ref{predmean} is the best linear unbiased predictor (BLUP) for $Y(X^*|X)$, the conditional distribution of the response at $X^*$ given data at $X$,  even when the normality assumption of the Gaussian process is violated. 
The $n$-dimensional row vectors of $K_\theta(X^*, X) K_\theta(X, X)^{-1}$ are referred to as the \emph{kriging weights}, and these are the vectors our method assumes are sparse for computational efficiency.
Taking the posterior mean prediction given in Equation~\ref{predmean} consists of computing the inner product of the kriging weights with the de-trended data vector $Y(X)$. 
Most covariances exhibit no general closed form solution to directly compute these weights without forming and inverting the matrix $K_\theta(X, X)$, which can be prohibitively expensive in large training data. 

By the properties of the GP assumption, the log-likelihood of the observations $Y(X)$ follow a jointly multivariate normal distribution:
\begin{equation}
	\label{ll}
	log(L(\theta, Y(X))) = - \frac{p}{2}log(2 \pi)  - \frac{1}{2} log(|K_\theta(X, X)|)  - \frac{1}{2} Y(X)^T K_\theta(X, X)^{-1} Y(X).
\end{equation}
Often $K_\theta(X, X)$ is assumed to be stationary and isotropic. 
That is, the covariance function is assumed to be a function of only the distances between the elements of $X$. 
Stationarity implies that for locations $\mathbf{x}_i$ and $\mathbf{x}_j$,
\begin{equation}
	K_\theta (\mathbf{x}_i, \mathbf{x}_j) = \phi_\theta(||\mathbf{x}_i - \mathbf{x}_j||_2),
\end{equation}
where $\phi_\theta$ is a functional form with parameters $\theta$. 
One of the most common covariance forms $\phi_\theta$ is the  Mat\' ern covariance function.
With covariance hyperparameters $\theta = \{\sigma^2, \rho, \nu, \tau^2\}$, and where $d$ is the distance between two locations in $X$,

\begin{equation} \label{eq:matern}
	\phi_\theta(d) = \sigma^2 \left[\frac{2^{1- \nu}}{\Gamma(\nu)} \Bigg( \sqrt{2\nu} \frac{d}{\rho} \Bigg)^{\nu} K_\nu \Bigg( \sqrt{2\nu} \frac{d}{\rho} \Bigg) + \tau^2 \mathbb{I} \{d=0\}\right],
\end{equation}
where $K_\nu$ is the modified Bessel function of the second kind. 
As $\nu \to \infty$, this form converges pointwise to the well-known Gaussian (RBF) covariance function.

Conventional approaches to training GP models estimate the covariance parameters using the log-likelihood in Equation \ref{ll} via maximum likelihood estimation, Bayesian analysis using Markov Chain Monte Carlo (MCMC), or by  way of grid search cross validation.
However, these estimation methods are too expensive to compute in large data since they require at least $O(n^3)$ computation and $O(n^2)$ memory.
Investigators have developed scalable approximate methods and models like those aforementioned in the introduction in order to perform such estimation in large datasets.
However, trends thus far in the literature indicate a trade-off between speed and accuracy: those methods with the fastest-time-to-solution on benchmark data yield posterior mean predictions that are not competitive in terms of RMSE with more expensive but accurate approximate methods \cite{heaton2019case}. 
We aim to obtain the best of both worlds: accurate stationary Gaussian process predictions that maintain best-in-class efficiency and scalability, while maintaining the accurate uncertainty quantification. 

\section{\texttt{MuyGPs}}
\label{sec:muygp}

We describe a novel approximate method for training stationary Gaussian process hyperparameters using leave-one-out cross-validation restricted to local predictions. 
Our methodology derives from the union of two insights: optimization by way of leave-one-out cross-validation allows us to avoid evaluating the likelihood in Equation~\ref{ll}, and restriction to the $k$ nearest neighbors of a prediction location limits the cost of computing the kriging weights $K_\theta(X^*, X) K_\theta(X, X)^{-1}$ in Equation~\ref{predmean} to $O(k^3)$.
While other investigators may have exploited both of these observations in different ways, our \texttt{MuyGPs} estimation method is the first to take advantage of both simultaneously to accelerate kernel hyperparameter estimation by enforcing sparsity in the kriging weights.

Leave-one-out cross-validation seeks hyperparameter values that minimize the sum of an out-of-sample loss associated with computing the posterior distribution of each training observation conditioned on all of the others.
To formally describe the procedure, let $\theta$ denote the hyperparameters that require estimation and $K_\theta(\cdot,\cdot)$ a GP kernel function of interest.
In the  Mat\'ern kernel, define $\theta=( \sigma^2 , \nu, \ell, \tau^2)^T$, but this method is applicable to other kernel forms.
Here, we fix $\sigma^2=1$ in estimation of the other parameters here since posterior mean predictions do not depend on overall scale parameter $\sigma^2$.
Hence, $\sigma^2$ is not estimable via any cross-validation method for all kernel forms.
We will introduce a different efficient estimation protocol for $\sigma^2$ at the end of this section after its fellow parameters have been estimated.

We must formulate the leave-one-out prediction of $Y(\mathbf{x}_i)$ given the set of all training points excluding $\mathbf{x}_i$. 
Define $X_{-i}=(\mathbf{x}_1, \mathbf{x}_2, \dots, \mathbf{x}_{i-1}, \mathbf{x}_{i+1}, \dots, \mathbf{x}_n)$ and $Y(X_{-i}) = (Y(\mathbf{x}_1), Y(\mathbf{x}_2), \dots, Y(\mathbf{x}_{i-1}), Y(\mathbf{x}_{i+1}), \dots, Y(\mathbf{x}_n))$ to be the training locations and observed responses excluding the $i$th training observation.
Then we modify Equation \ref{predmean}  
to obtain the mean GP prediction by regressing the training labels on the corresponding inputs,
\begin{equation}
	\label{cvpred1}
	\widehat{Y}(\mathbf{x}_i \mid X_{-i}) = K_\theta(\mathbf{x}_i, X_{-i}) K_\theta(X_{-i}, X_{-i})^{-1} Y(X_{-i}).
\end{equation}
Here $K_\theta(\mathbf{x}_i, X_{-i})$ is the cross-covariance between $\mathbf{x}_i$ and $X_{-i}$, while $K_\theta(X_{-i}, X_{-i})$ is the covariance among points in $X_{-i}$, both in terms of $K_\theta(\cdot, \cdot)$. 
Although many criterion for predictions accuracy are possible, we select the mean squared error criterion
\begin{equation} \label{eq:loss}
	Q(\theta) = \frac{1}{n} \sum_{i=1}^{n} \left ( Y(\mathbf{x}_i) - \widehat{Y}(\mathbf{x}_i  \mid X_{-i}) \right )^2.
\end{equation}
The hyperparameters are estimated to minimize the leave-one-out cross-validation loss
\begin{equation} \label{eq:objective}
	\widehat{\theta}=\min_{\theta} Q(\theta).
\end{equation}
In our experiments we utilize the L-BFGS-B algorithm \cite{zhu1997algorithm} to minimize Equation~\ref{eq:objective}.

When there are a large number of observations, realizing an instance of the cross-validation loss in Equation~\ref{eq:loss} so described is even more computationally expensive than the loglikelihood in Equation~\ref{ll};
the procedures have complexity $O(n^4)$ and $O(n^3)$, respectively.
This suggests that the optimization given by Equation~\ref{eq:objective} is even less efficient than traditional maximum likelihood estimation.
In order to achieve our scalability objectives, we replace the full kriging of Equation~\ref{cvpred1} with \emph{local kriging}, using the $k$ nearest neighbor locations of $\mathbf{x}_i$ instead of all of the $n-1$ locations denoted by $X_{-i}$.
This modifies the complexity of leave-one-out cross-validation (computing $Q(\theta)$) to $O(n k^3)$, which is much more scalable than likelihood-based approaches when $k \ll n$. 

We now illustrate our method precisely.
Let $N_i$ be the set of $k$ indices in $\{1, \dots, i-1, i+1, \dots, n\}$ indicating those elements of $X_{-i}$ that are nearest to $\mathbf{x}_i$ in terms of distance.
Similarly, define $X_{N_i}$ as the set of training observations nearest to $\mathbf{x}_i$, and let $Y(X_{N_i})$ be their corresponding responses.
This allows us to modify our leave-one-out prediction in Equation~\ref{cvpred1} to
%
\begin{equation}
	\label{cvpred2}
	\widehat{Y}_{NN}(\mathbf{x}_i \mid X_{N_i}) = K_\theta(\mathbf{x}_i, X_{N_i}) K_\theta(X_{N_i}, X_{N_i})^{-1} Y(X_{N_i}),
\end{equation}
where the kriging weights and observed responses are defined in terms of $N_i$.

Modifying Equation~\ref{eq:loss} in terms of the nearest neighbor leave-one-out predictions from Equation~\ref{cvpred2}, the resulting objective function requires $O(n k^3)$ operations to evaluate.
We can further reduce our training complexity overhead of $Q(\theta)$ by utilizing batching, a common technique in machine learning. 
Let $B$ be a subset of $b$ indices randomly sampled without replacement from $\{1, \dots, n\}$. 
We can then modify Equation~\ref{eq:loss} by summing only over the nearest neighbor leave-one-out squared error of locations $\mathbf{x}_i$ such that $i \in B$, whereby we obtain the modified loss function
\begin{equation} \label{eq:batch_loss}
	Q_{B}(\theta) = \frac{1}{b} \sum_{i \in B} \left ( Y(\mathbf{x}_i) - \widehat{Y}_{NN}(\mathbf{x}_i \mid X_{N_i}) \right )^2.
\end{equation}

This approximation introduces some additional variability into the optimization problem since the batch indices are randomly selected, but the computational savings can be significant if $b \ll n$, bringing the cost of evaluating the loss function to $O(bk^3)$, which importantly is independent of $n$.
It is also important to note that the nearest neighbor index sets $N_i$ utilized in the evaluations of $\widehat{Y}_{NN}(\mathbf{x}_i)$ for $i \in B$ in Equation~\ref{eq:batch_loss} still range over the full $n$ data points $X$, rather than only those data indicated by $B$.
If nearest neighbor candidates were restricted to batched points only, then the nearest neighbors of each batched observation would be artificially distant from one another.
Hence, more data is ultimately used to estimate the hyperparameters than is included directly in the batch.

Although $\sigma^2$ is not needed for the mean predictions, the prediction errors depend linearly on its estimation since for batched observations it is defined as
\begin{equation} \label{cverr}
	\text{Var}(\widehat{Y}_{NN}(\mathbf{x}_i \mid X_{N_i})) = K_\theta(\mathbf{x}_i, \mathbf{x}_i) - K_\theta(\mathbf{x}_i, X_{N_i}) K_\theta(X_{N_i}, X_{N_i})^{-1} K_\theta(X_{N_i}, \mathbf{x}_i) .
\end{equation}
After all other parameters are estimated via minimization of \eqref{eq:batch_loss}, we obtain $\widehat{\sigma^2}$. 
Define $\Omega_\theta= \frac{K_\theta}{\sigma^2}$. 
In general, there is a closed fom equation for the maximum likelihood estimate for $\sigma^2$ given $\theta$.
However, this closed form solution involves forming and inverting the full training covariance matrix $K_\theta(X, X)$, which is not possible for large $n$.
Our approximate estimate is the mean scale parameter within each batched $k$ neighborhood or precisely
\begin{equation}
	\widehat{\sigma^2} = \frac{1}{kb} \sum_{i \in B} Y(X_{N_i})^T \Omega_\theta(X_{N_i}, X_{N_i})^{-1} Y(X_{N_i}).
\end{equation}

Finally, predictions at testing (prediction) locations are computed using a similarly local design. 
Define $X_{N_i^{\star}}$ as the set of training observations nearest to the $i$th testing location $\mathbf{x}_i^{\star}$, and let $Y(X_{N_i^{\star}})$ be their corresponding responses.
Then, approximate  predictions and prediction errors are obtained via \texttt{MuyGPs} as

\begin{align}
	\widehat{Y}(X^* \mid X) &\approx K_{\hat{\theta}}(X^*, X_{N_i^{\star}}) K_{\hat{\theta}}(X_{N_i^{\star}}, X_{N_i^{\star}})^{-1} Y(X_{N_i^{\star}}), \text{ and}
	\label{predmean2}\\
	\text{Var}(\widehat{Y}(X^* \mid X))  &\approx K_{\hat{\theta}}(X^*, X^*) - K_{\hat{\theta}}(X^*, X_{N_i^{\star}}) K_{\hat{\theta}}(X_{N_i^{\star}}, X_{N_i^{\star}})^{-1} K_{\hat{\theta}}(X_{N_i^{\star}}, X^*),
	\label{predvar2}
\end{align}
where $\hat{\theta}$ are the covariance parameters trained as aforementioned.

\section{Numerical Studies}
\label{sec:sim}

We demonstrate the effectiveness of our novel GP hyperparameter estimation on a benchmark temperature dataset.
The dataset considered is sourced from \cite{heaton2019case}, is pictured in Figure \ref{fig:data}, and is available for download at  https://github.com/finnlindgren/heatoncomparison.
The data were collected using the Terra instrument onboard the MODIS satellite.
It is composed of daytime land surface temperatures for August 4, 2016 in longitudes -95.91153 to -91.28381 and latitudes 34.29519 to 37.06811 observed on a $500 \times 300$ grid.
This region is attractive due to its completeness, as 98.9\%  of locations were observed in the area of interest on August 4, 2016.
The cloud coverage pattern from August 6, 2016 was applied to the data in order to create a realistic missingness pattern for the training/testing data. 
This yielded 105,569 training observations, 42,740 testing observations, and 1,691 missing (non-testing) observations, which are excluded from analysis. 
We normalize the gridded observations to be in $[0,1]^2$ while maintaining the relationship among latitude and longitude by
$$ \mathbf{x}_i = \frac{(latitude_i, longitutde_i)+218}{464},$$
for $i=1,2,...105,569$ for all observations given in the training dataset. 
This transformation creates a range of (0.2102833, 0.8078952) in latitude and (0.001297902, 0.998651208) in longitude in the training observations.
We apply this same normalizing transformation to the testing locations.

We compute the same statistics to evaluate our methods used in the competition in \cite{heaton2019case}. 
We consider mean absolute error (MAE), root mean squared error (RMSE), continuous rank probability score (CRPS) \cite{gneiting2007strictly}, interval score (INT) \cite{gneiting2007strictly}, and empirical coverage of 95\% confidence intervals (COV). 
We computed CRPS and INT source code drawn directly from \cite{heaton2019case} for consistency.
The timings of our method were obtained with a 2016 MacBook Pro with 2.9 CHz Quad-Core Intel Core i7 with 16 GB of RAM.
In \cite{heaton2019case}, a machine with 256 GB of RAM and 2 Intel Xeon E5-2680 v4 2.40GHz CPUs with 14 coreseach and 2 threads per core - totaling 56 possible threads for use in parallel computing was utilized for comparison.

We implemented our newly-developed \texttt{MuyGPs} methodology as described in Section \ref{sec:muygp}.
Prior to fitting we must make modeling decisions including the selection of a mean function and covariance function, the choice of batch size, and the number of nearest neighbors to be employed.
We realized models using a variety of different choices in order to demonstrate the robustness of our results.
First, we assumed that the response function $Y$ has mean zero throughout the entirety of this manuscript.
Although this is a common assumption when fitting stationary GPs, it is conventional and necessary in real-world data to first remove mean trends prior to GP fitting. 
Therefore, define $Y^{obs}$ to be the vector of non-zero mean temperature responses,
and $Y_k(x) = Y^{obs}(X) - \mu_k(X)$, where $Y_k(X)$ is the mean zero response with mean function $\mu_k(X)$ removed (previously notated $Y(X)$). 
We consider three simple mean functions so that $k=1,2,3$.

First, we consider a constant mean function. Let
$$ \mu_1(X) = c.$$
This simple mean is estimated by the sample mean of the training data \\
$\hat{c}=\frac{1}{n} \sum_{i=1}^nY^{obs}(\mathbf{x}_i)$. 
Next, we consider a linear mean function with an interaction term.
Define $Z$ to be the $n \times 4$ matrix that contains a $n\times 1$ vector of ones, $X$, and the multiplication of the rows elements of $X$ so that  $Z=[1_n, X^T, X^T[,1]*X^T[,2]]$. Then define
$$\mu_2(X)=Z\beta,$$
where $\beta$ is a $4 \times 1$ vector of mean parameters.
These parameters are estimated using the ordinary least squares solution of the training data.

\begin{figure}
	\centering
	\begin{subfigure}[b]{0.49\textwidth}
		\centering
		\includegraphics[width=\textwidth]{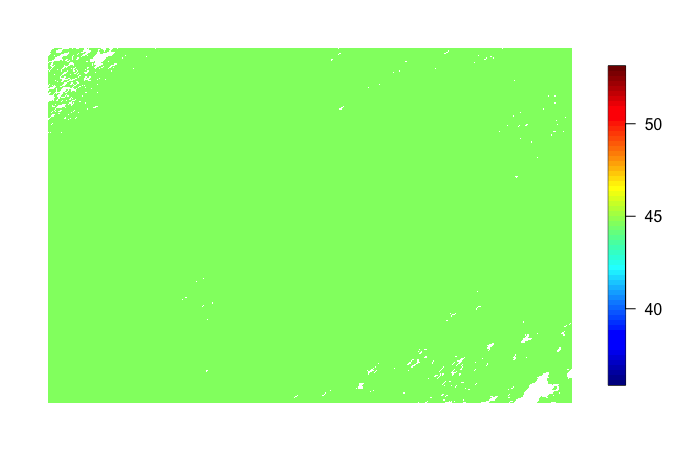}
		\caption{$\mu_1$}
		\label{fig:mu1}
	\end{subfigure}
	\hfill
	\begin{subfigure}[b]{0.49\textwidth}
		\centering
		\includegraphics[width=\textwidth]{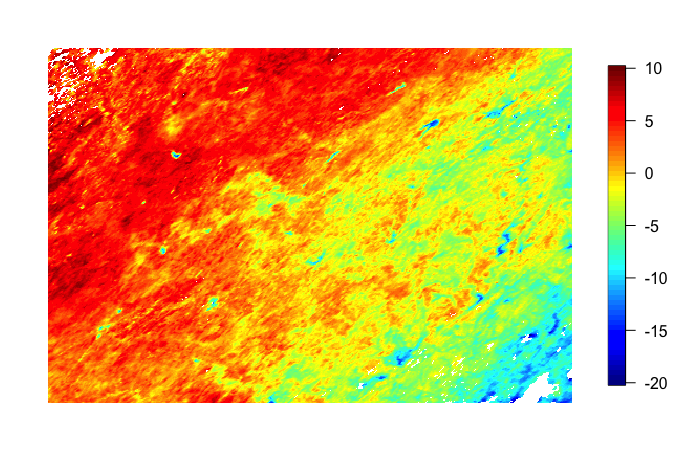}
		\caption{$Y_1$}
		\label{fig:y1}
	\end{subfigure}
	\hfill
		\begin{subfigure}[b]{0.49\textwidth}
		\centering
		\includegraphics[width=\textwidth]{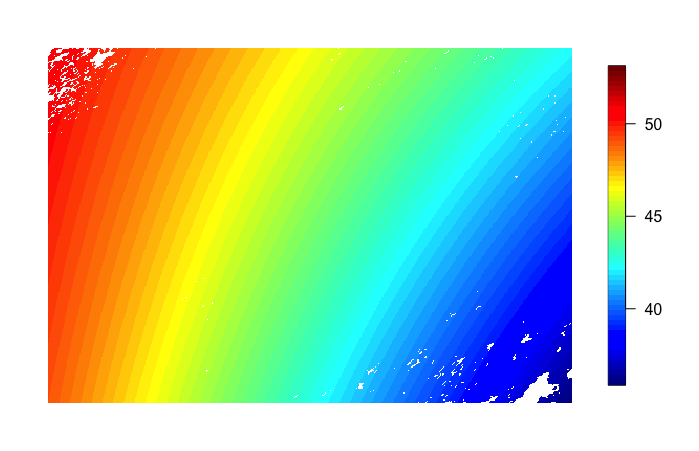}
		\caption{$\mu_2$}
		\label{fig:mu2}
	\end{subfigure}
	\hfill
	\begin{subfigure}[b]{0.49\textwidth}
		\centering
		\includegraphics[width=\textwidth]{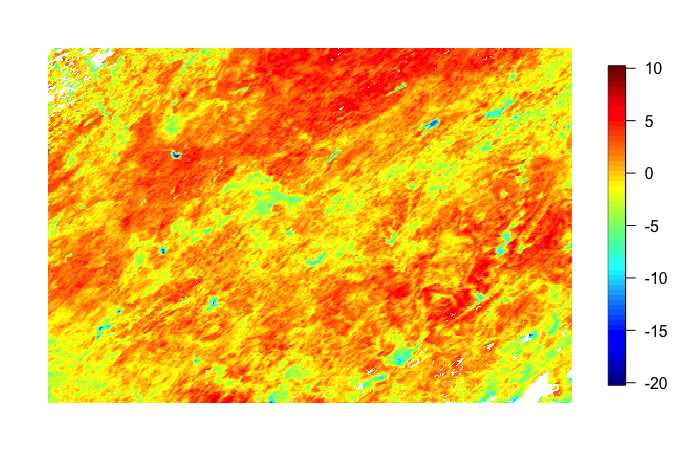}
		\caption{$Y_2$}
		\label{fig:y2}
	\end{subfigure}
\hfill	\begin{subfigure}[b]{0.49\textwidth}
	\centering
	\includegraphics[width=\textwidth]{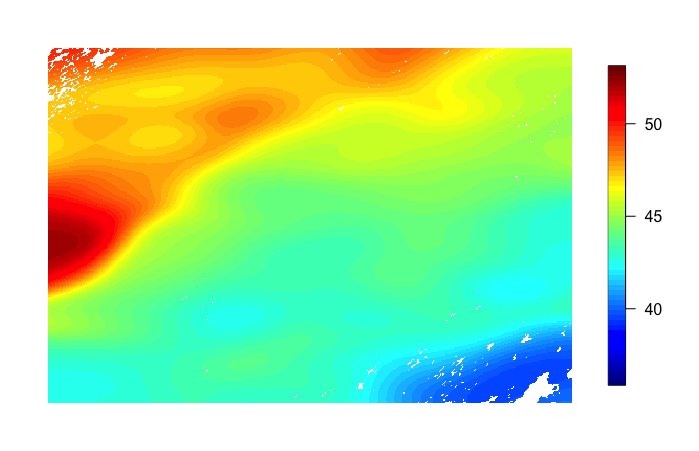}
	\caption{$\mu_3$}
	\label{fig:mu3}
\end{subfigure}
\hfill
\begin{subfigure}[b]{0.49\textwidth}
	\centering
	\includegraphics[width=\textwidth]{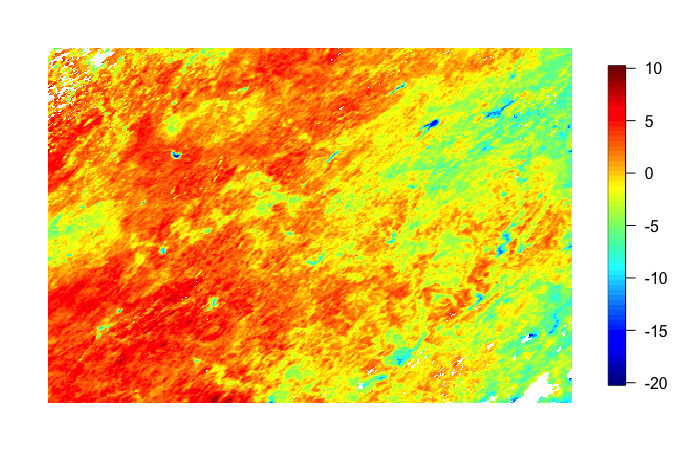}
	\caption{$Y_3$}
	\label{fig:y3}
\end{subfigure}
	\caption{Mean functions considered, and the residual observations that are fit via \texttt{MuyGPs} using the testing data. }
	\label{fig:mean}
\end{figure}

Finally, we consider a mean that is based on kernel smoothing. 
In particular, we utilize the Nadaraya-Watson kernel smoother \cite{nadaraya1964estimating,watson1964smooth}. 
Define
$$\mu_3(X)= \frac{\sum_{i=1}^nG(X,\mathbf{x}_i)Y(\mathbf{x}_i)}{\sum_{i=1}^n G(X,\mathbf{x}_i)},,$$
where $G(X,\mathbf{x}_i)$ is the Gaussian kernel observed for locations $X$ and $\mathbf{x}_i$ \\ ($\exp{-||X-\mathbf{x}_i||/\rho}$).
We use the $smooth.2d$ function from the package $fields$ in order to estimate implement this smoothing with length scale $\rho=25$ on the $500 \times 300$ grid \cite{fields}. This implementation utilizes a 2-D fast Fourier transform (FFT) to efficiently estimate the smoother using the partially gridded training data \cite{fields}.
Employing this mean is similar to the multi-resolution principle in \cite{nychka2015multiresolution,katzfuss2017multi}, where very smooth low frequency trends are fit by this filtering mean, and the high-frequency trend are fit on resulting residuals.
We plot the three mean functions ($\mu_k$ ), and the corresponding data fit in each mean ($Y_k$) is seen  in Figure \ref{fig:mean}.
Note that because $\mu_2(X)$ and $\mu_3(X)$ depend on $X$, they lead to first order non-stationary predictions.

\texttt{MuyGPs} can theoretically estimate the parameters of most kernel functions.
For its flexibility, we employ the Mat\'ern covariance as stated in Equation \ref{eq:matern} in our experiments.
Since the impact on prediction is low, we fix $\tau^2=0.001$, and if otherwise not stated, we fix the length scale $\rho = 1.0$. 
In our experiments, $\nu$ and $\rho$ are difficult to simultaneously estimate, and we found better performance when $\nu$ is allowed to vary. 
Further, we explore other fixed values of $\rho$ in the results in Table \ref{tab:perf}.

\begin{figure}
	\centering
	\includegraphics[width=.6\linewidth]{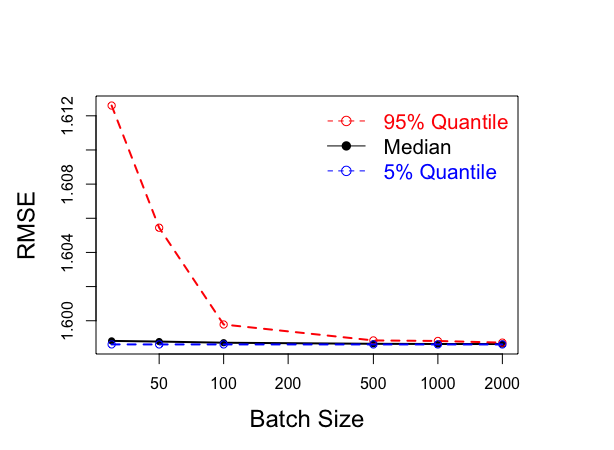}
	\caption{Empirical 90\% confidence intervals of RMSE over 100 simulation iterations for 50 exact nearest neighbors with $\mu_3$.}
	\label{fig:batch}
\end{figure}

\begin{figure}
	\includegraphics[width=\linewidth]{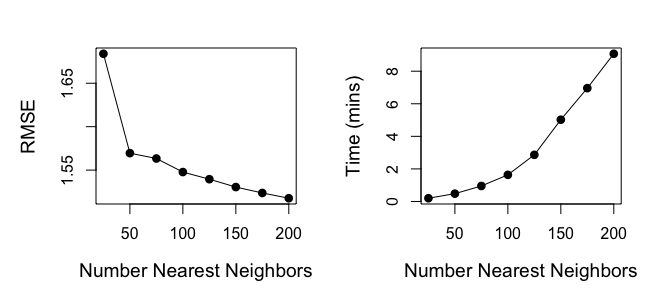}
	\caption{Performance in RMSE and computing time for 500 batched values using approximate nearest neighbors and $\mu_3$.}
	\label{fig:nn}
\end{figure}

Next, we must select a batch size for the \texttt{MuyGPs} estimation procedure.
In all of our experiments we sampled batch elements from our training data randomly and without replacement. 
Using exact nearest neighbor sets of size 50 and $\mu_3$, we demonstrate the variability in prediction performance using RMSE for various batch sizes in Figure~\ref{fig:batch} for 100 independent simulation iterations at each batch size.
We then compute empirical 95\% empirical confidence intervals at each batch size to determine the variability of estimates from that batch size.
In even extremely small batch sizes of 25 members, the variability in the RMSE is very small at a difference of about 0.014 length in a 95\% empirical confidence interval of the simulations.
When considering batch sizes of 500, there is less than 0.001 in variability, and at batch sizes of 2,000, there is nearly no variability in prediction RMSE.
Note that this dataset has 105,569 training observations, so a batch size of 2,000 offers significant improvement where we can use less that 1.8\% of the data in the leave-one-out cross-validation optimization without impacting the results.

\begin{figure}
	\includegraphics[width=\linewidth]{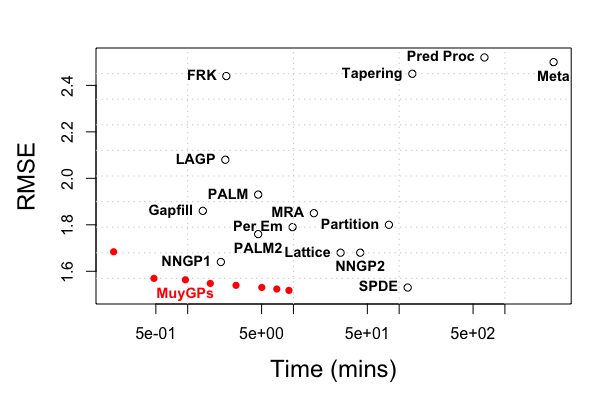}
	\caption{Comparison  in RMSE and computing time in comparison of \texttt{MuyGPs} to all methods in \cite{heaton2019case} and \cite{edwards2020precision}. 
		The multiple \texttt{MuyGPs} observations are the results from Figure \ref{fig:nn} with varying nearest neighbor set sizes between 25 and 200 in 25 incriments.
	}
	\label{fig:compare}
\end{figure}

Following the tradeoffs associated with batch size selection, we demonstrate selecting the number of nearest neighbors. 
As more neighbors are incorporated into the model, the predictions become more accurate, but the computing time is increased. 
We can achieve practical speed improvements by employing approximate nearest neighbor algorithms such as the hierarchical navigable small world (HNSW) algorithm \cite{malkov2018efficient}.
HNSW offers further scaling benefits when the number of variables in $\mathbf{x}_i$ is large.
We implement this via bindings for the 'hnswlib' library available at https://github.com/nmslib/hnswlib.
We demonstrate this performance tradeoff in including different numbers of approximate nearest neighbors in Figure \ref{fig:nn} using $\mu_3$ as an example.
There is a large accuracy performance gap between 25 and 50 nearest neighbors, but all increases in number of nearest neighbors improves the prediction accuracy, but increases total computing time due to the $O(k^3)$ term associated with solving the nearest neighbor systems of linear equations.

\begin{table}[ht]
	\centering
	\caption{Numerical scoring for various versions of \texttt{MuyGPs} with competing method original results from \cite{heaton2019case} and \cite{edwards2020precision} for comparison.}
	\begin{tabular}{rrrrrrrr}
		Method& $\ell $ & MAE & RMSE  & CRPS& INT & COV & Time (min)\\ 
		\hline
		&0.1& 1.14 & 1.66 & 0.86 & 9.25 & 0.95 & 0.70 \\ 
		& 0.25  & 1.15 & 1.64 & 0.84 & 8.40 & 0.95 & 0.68 \\ 
		\texttt{MuyGPs}, $\mu_1$ & 0.5 & 1.19 & 1.67 & 0.85 & 8.02 & 0.93 & 0.75 \\ 
		&0.75 & 1.21 & 1.69 & 0.86 & 7.90 & 0.93 & 0.75 \\ 
		& 1.0 & 1.22 & 1.68 & 0.86 & 7.85 & 0.92 & 0.63 \\ 
		\hline
		& 0.1  & 1.12 & 1.64 & 0.86 & 9.38 & 0.95 & 0.64 \\ 
		& 0.25 & 1.13 & 1.62 & 0.83 & 8.31 & 0.94 & 0.64 \\ 
		\texttt{MuyGPs}, $\mu_2$ & 0.5 & 1.15 & 1.62 & 0.83 & 8.00 & 0.94 & 0.64 \\ 
		&0.75 & 1.19 & 1.65 & 0.85 & 7.85 & 0.93 & 0.63 \\ 
		& 1.0 & 1.19 & 1.64 & 0.84 & 7.80 & 0.93 & 0.63 \\ 
		\hline
		& 0.1 & 1.07 & 1.54 & 0.84 & 9.35 & 0.95 & 0.76 \\ 
		& 0.25 & 1.08 & 1.53 & 0.80 & 8.24 & 0.94 & 0.66 \\ 
		\texttt{MuyGPs}, $\mu_3$ & 0.5 & 1.12 & 1.55 & 0.81 & 7.95 & 0.94 & 0.62 \\ 
		& 0.75 & 1.14 & 1.56 & 0.81 & 7.78 & 0.93 & 0.63 \\ 
		& 1.0 & 1.15 & 1.57 & 0.82 & 7.71 & 0.93 & 0.65 \\ 
		\hline
		\hline
		FRK& &1.96 & 2.44 & 1.44 & 14.08 & 0.79 & 2.32 \\
		Gapfill && 1.33 & 1.86 & 1.17 & 34.78 & 0.36 & 1.39 \\
		LatticeKrig& &1.22 & 1.68 & 0.87 & 7.55 & 0.96 & 27.92 \\
		LAGP&& 1.65 & 2.08 & 1.17 & 10.81 & 0.83 & 2.27 \\
		Metakriging&& 2.08 & 2.50 & 1.44 & 10.77 & 0.89 & 2888.52 \\
		MRA&& 1.33 & 1.85 & 0.94 & 8.00 & 0.92 & 15.61 \\
		NNGP& & 1.21 & 1.64 & 0.85 & 7.57 & 0.95 & 2.06 \\
		NNGP2& & 1.24 & 1.68 & 0.87 & 7.50 & 0.94 & 42.85 \\
		Partition && 1.41 & 1.80 & 1.02 & 10.49 & 0.86 & 79.98 \\
		Pred. Proc. && 2.15 & 2.64 & 1.55 & 15.51 & 0.83 & 160.24 \\
		SPDE & & 1.10 & 1.53 & 0.83 & 8.85 & 0.97 & 120.33 \\
		Tapering& &1.87 & 2.45 & 1.32 & 10.31 & 0.93 & 133.26 \\
		Per Embed& &1.29 & 1.79 & 0.91  & 7.44 & 0.93 & 9.81 \\
		\hline
		\hline
		PALM	& &1.59 & 1.93 & 1.15 & 11.78 & 0.78 & 4.64\\
		Global + PALM2	& &1.44& 1.76 & 1.03  & 9.28 & 0.84 & 4.64 \\
	\end{tabular}
	\label{tab:perf}
\end{table}

Finally, we compare the performance of our method to all previous methods computed on this data in \cite{heaton2019case} and \cite{edwards2020precision}. 
We plot the tradeoffs between RMSE and time visually in Figure \ref{fig:compare}, where we consider multiple numbers of nearest neighbors in \ref{fig:nn}, with $\mu_3$ , 500 batched values, and $\rho=1$.
Notably, \texttt{MuyGPs} with a small number of nearest neighbors is the fastest method presented and remains among the top five methods in terms of RMSE. 
Additionally, \texttt{MuyGPs} with a large number of nearest neighbors is the most accurate method considered in all manuscripts.
Further, \texttt{MuyGPs} with number of nearest neighbors greater than or equal to 50 ranks 2nd or better in all the methods in accuracy.
In \cite{heaton2019case}, SPDE \cite{lindgren2011explicit} is the most accurate method, but it took approximately 120 minutes to compute, which is roughly more than 12 times slower than the slowest presented \texttt{MuyGPs} setting.
A similarly accurate solution is obtained by \texttt{MuyGPs} in approximately only 5 minutes.

We consider the other statistics based on uncertainty in the results in Table \ref{tab:perf}.
In all previous studies, the parameter $\ell$ from the Mat\'ern covariance has been fixed at 1.0.
This is because the $\ell$ and $\nu$ parameters are unidentifiable to estimate simultaneously. 
Given this were a blind competition as the participants of \cite{heaton2019case}, this value would likely have been fixed at a reasonable value. 
To show the robustness of our method, we demonstrate the performance of \texttt{MuyGPs} for several selections of fixed $\ell$ values. 
In fact, our previous selection of $\ell=1.0 $ is the least favorable reasonable value for $\ell$ in the main results of Table \ref{tab:perf}.
With all three of the mean functions, even with only 50 approximate nearest neighbors, the accuracy of \texttt{MuyGPs} is competitive.
Further, all methods compared were under a minute in computing time, which is faster than any method previously tested on this data.
In all cases, the coverage is near to that of the the expected 0.95, and the other uncertainty interval statistics perform favorable in the field of methods.
In summary, the \texttt{MuyGPs} method offers more accurate and faster computation time than any existing method, and no matter the modeling choices, is a competitive with all known scalable GP computation methods.

\section{Discussion}
\label{sec:discuss}

We have presented in this manuscript \texttt{MuyGPs}, a new method for stationary Gaussian process hyperparameter estimation and prediction that is both computationally efficient and highly performant when compared to state-of-the-art approximate GP estimation methods. 
\texttt{MuyGPs} minimizes the leave-one-out cross-validated mean squared error using only data from the $k$ nearest neighbors for selected locations across the domain.
Similarly, predictions depend only upon their $k$ nearest neighbor training observations. 
We demonstrate that although this method is simple, it is powerful and performs well against state-of-the-art approximation methods. 

Using this method requires the selection of a mean function, covariance function, batch size, and number of nearest neighbors. 
We demonstrate that our method performs well with very few batched points.
In fact, when exact nearest neighbors are employed, the accuracy has nearly no variance with a batch size of about 2,000, which is approximately 1.8\% the training data.
Increasing the number of nearest neighbors improves performance in terms of RMSE at the cost of increasing computing time.
Finally, we show that a non-stationary mean function improves the accuracy of our model, but the magnitude of this effect is small compared to changes in GP estimation methods. 
Although these selections ultimately define a tradeoff between performance in accuracy and computation time, our method performs favorably compared to existing methods regardless of these modeling choices.
Settings of our method offers both the fastest, and the most accurate method among all known methods, and performs among the top few methods in all other considered statistics.


To date we have optimized our method only with respect to MSE, but other objective functions are possible. 
Using optimal parameters for this testing dataset yields an RMSE of 1.42 with 0.95 coverage via our model with only 50 approximate nearest neighbors (although these are not the parameters selected by our training method).
This implies there may be room for improvement of parameter estimation using a more complex objective function.
One possible idea could be to incorporate other statistics from Table \ref{tab:perf} into the objective function.

Although we achieve a first-order non-stationary predictions through \\non-stationary mean functions, our hyperparameter estimation method is amenable to estimation of second-order non-stationary models that have previously been considered too expensive for large data.
For example, one could assume hierarchically that, for example, $\nu$ parameter follows a GP over the domain. 
Although we found small batch sizes relatively effective with a random sample under a stationarity assumption, more care may be needed in order to select the batched observations and their size under this more complex non-stationary model case.


\section*{Acknowledgments}
This work was performed under the auspices of the U.S. Department of Energy by Lawrence Livermore National Laboratory under Contract DE-AC52-07NA27344 with IM release number LLNL-JRNL-822013.
Funding for this work was provided by LLNL Laboratory Directed Research and Development grant 19-SI-004.

This document was prepared as an account of work sponsored by an agency of the United States government. Neither the United States government nor Lawrence Livermore National Security, LLC, nor any of their employees makes any warranty, expressed or implied, or assumes any legal liability or responsibility for the accuracy, completeness, or usefulness of any information, apparatus, product, or process disclosed, or represents that its use would not infringe privately owned rights. Reference herein to any specific commercial product, process, or service by trade name, trademark, manufacturer, or otherwise does not necessarily constitute or imply its endorsement, recommendation, or favoring by the United States government or Lawrence Livermore National Security, LLC. The views and opinions of authors expressed herein do not necessarily state or reflect those of the United States government or Lawrence Livermore National Security, LLC, and shall not be used for advertising or product endorsement  purposes.

\bibliographystyle{siamplain}
\bibliography{bib}
\end{document}